\documentclass{emulateapj}
\usepackage[below]{placeins}
\usepackage{natbib}
\slugcomment{To be published by the Astrophysical Journal Letters}

\def\gtorder{\mathrel{\raise.3ex\hbox{$>$}\mkern-14mu
    \lower0.6ex\hbox{$\sim$}}}
\def\ltorder{\mathrel{\raise.3ex\hbox{$<$}\mkern-14mu
    \lower0.6ex\hbox{$\sim$}}}

\shorttitle{Self-Gravitating Accretion Flows}
\shortauthors{Begelman \& Shlosman}

\begin{document}

\title{Angular Momentum Transfer and Lack of Fragmentation\\ in 
Self-Gravitating Accretion Flows}

\author{ 
Mitchell C. Begelman\altaffilmark{1} and Isaac Shlosman\altaffilmark{2}
}
\affil{JILA, 440 UCB, 
     University of Colorado, \\ 
     Boulder, CO 80309-0440, USA \\ 
     email: {\tt mitch@jila.colorado.edu; shlosman@pa.uky.edu}}

\altaffiltext{1}{Also at Department of Astrophysical and Planetary Sciences, University of Colorado at Boulder}
\altaffiltext{2}{JILA Visiting Fellow}

\begin{abstract}
Rapid inflows associated with early galaxy formation lead to the accumulation of
self-gravitating gas in the centers of proto-galaxies. Such gas accumulations
are prone to non-axisymmetric instabilities, as in the well-known Maclaurin sequence
of rotating ellipsoids, which are accompanied by a catastrophic loss of angular
momentum ($J$). Self-gravitating gas is also intuitively associated with star formation.
However, recent simulations of the infall process display highly turbulent
continuous flows. We propose that $J$-transfer,
which enables the inflow, also suppresses fragmentation.  Inefficient
$J$ loss by the gas leads to decay of turbulence, 
triggering global instabilities and renewed turbulence driving.
Flow regulated in this way is stable against fragmentation, whilst 
staying close to the instability threshold for bar formation --- thick 
self-gravitating disks are prone to global instabilities before they become unstable
locally. On smaller scales, the fraction of gravitationally
unstable matter swept up by shocks in such a flow is a small and decreasing function 
of the Mach number.  We conclude counterintuitively that gas able to 
cool down to a small fraction of its virial temperature will not fragment as it 
collapses.  This provides a venue for supermassive black holes to form via direct infall,
without the intermediary stage of forming a star cluster. Some
black holes could have formed or grown in massive
halos at low redshifts. Thus
the fragmentation is intimately related to $J$
redistribution within the system: it is less dependent on the 
molecular/metal cooling but is conditioned by the ability of the flow to develop
virial, supersonic turbulence.  
\end{abstract}

    
%-------------------------------------------------------------------
\section{Introduction}
\label{sec:intro}

The formation of luminous galactic nuclei and supermassive black holes is inherently 
related to the gravitational collapse of gas and its concurrent release of large 
quantities of angular momentum.  Recent numerical simulations of this process (Regan 
\& Haehnelt 2009; Levine et al. 2008; Wise, Turk \& Abel 2008; Englmaier \& Shlosman 
2004) 
reveal a number of puzzling empirical details. First, the collapse appears to be 
self-similar, over as many as 12 decades in radius, with a density profile of 
$\rho\propto r^{-2}$ when averaged over spheres, where $r$ is the spherical radius.
This is observed despite an extremely asymmetric and chaotic appearance locally.  
Second, it is accompanied by 
fully developed, highly supersonic turbulence, with near-virial velocities at all 
radii. And third, it does not show any fragmentation, despite being self-gravitating 
and isothermal. The gas temperature is stable at a few$\times 10^3$~K (sound speed
$c_{\rm s}\sim 3~{\rm km~s^{-1}}$), as governed by the atomic cooling rate, and the
flow Mach number is $\mathcal{M}\sim $~few. Here we propose an explanation for these 
phenomena.

Angular momentum transfer is of paramount importance to the evolution of many 
astrophysical systems.  Such systems, which are kept intact by their internal 
gravitational energy, or self-gravity, are found on all scales, from 
planetary and stellar, to galactic and beyond.  It is understood that (long-range) 
gravitational torques rather than (local) viscous torques are responsible for 
angular momentum redistribution in these systems (e.g., Lynden-Bell \& Kalnajs 1972), 
especially on scales where torques generated by ordered 
magnetic fields (Blandford \& Payne 1982) are inefficient.

The amplitude of gravitational torques in a gaseous self-gravitating object is 
determined by the gravitational quadrupole moment, i.e., by its departure from axial 
symmetry. When such an object is supported by rotation against gravitational 
collapse, its axial symmetry is known to be {\it spontaneously} broken when the 
ratio of bulk kinetic energy to (absolute) gravitational potential energy, $T/|W|$, 
is larger than some critical value. The Maclaurin sequence of rotating fluid 
ellipsoids, being an angular momentum, $J$, sequence, can serve as a simple 
quantitative example (see, e.g., Chandrasekhar [1969], for a review). At some value of 
$J$, the axial symmetry of an ellipsoid is broken abruptly by the $m=2$ mode to a 
barlike shape, due to the lower energy of this configuration. A more robust 
stability indicator, which is applicable to differentially rotating, centrally
concentrated fluids, is $\alpha\equiv (T/|W|)/(\Omega/\Omega_{\rm J})$, where 
$0\leq\alpha\leq 1/2$,
$\Omega$ is the angular velocity and $\Omega_{\rm J}=(2\pi G\rho A)^{1/2}$ 
is the Jeans frequency, $A$ being the shape factor defined in terms of the 
meridional eccentricity
(Christodoulou, Shlosman \& Tohline 1995). The bar instability appears to be 
universal and is 
triggered in any self-gravitating system, collisional or collisionless, with various 
mass density profiles and sufficient bulk rotation. The subsequent evolution, 
however, depends on the ability of the system to dissipate energy.

When self-gravitating systems shed angular momentum, their potential (i.e., 
self-gravitating) and total energies can decrease as well.  The relative amount of 
angular momentum loss, compared to potential energy loss, is crucial to determining 
the outcome. If this ratio is low enough, there is again a tendency to break axial 
symmetry and increase the quadrupole moment, as in the Maclaurin sequence.  In 
non-dissipative systems, of course, the total energy is conserved.  Such systems 
cannot be driven into a state of gravitational collapse by the loss of angular 
momentum, because the loss of potential energy saturates. On the other hand, in 
dissipative systems, energy loss can keep pace with the loss of angular momentum, 
forcing them into continuous collapse.  The latter situation applies in fluid 
systems, such as gas-dominated disks, and forms the essence of the `bars-in-bars' 
mechanism of angular momentum transfer in nested stellar/fluid or fluid/fluid bars, 
which is accompanied by dynamical inflow (Shlosman, Frank \& Begelman 1989; 
Shlosman, Begelman \& Frank 1990). Such fluid bars can form in maximally rotating 
neutron stars (e.g., Shibata, Baumgarte \& Shapiro 2000), trigger the formation of 
black holes (Begelman, Volonteri \& Rees 2006), or dominate galactic 
proto-disks at high and intermediate redshifts (e.g., Heller, Shlosman \&
Athanassouala 2007; Romano-Diaz et al. 2008), and hence provide a universal 
channel for angular momentum redistribution in self-gravitating systems.

\section{Turbulence driving}

To ensure continuous collapse, a rotation-dominated self-gravitating system must 
efficiently decrease its $J$ and remain dissipative. Stellar systems 
(i.e., disk galaxies) retain the first property, but plainly lack the second. What 
is less clear is whether gasdynamical systems are guaranteed to remain dissipative 
as they collapse.  Being dominated by self-gravity, they can fragment and these 
fragments can contract, sharply decreasing their interaction cross-sections and 
becoming essentially collisionless.  Star formation is an example of this 
process. The internal dissipation will be dramatically curtailed by the star 
formation and, consequently, the overall collapse of the system will be 
terminated or at least delayed. However, the ability of fragments to contract 
depends on their equation of state and should not be taken for granted (Paczy\'nski 
1978; Shlosman et al. 1990). The requirement that inflow accompanies 
the transfer of $J$ in nested fluid bars critically depends on the 
non-fragmentation of the flow. It has never been shown that the gasdynamical 
bars-in-bars mechanism can overcome this limitation.

Feedback from stellar evolution, e.g., from stars formed prior to the inflows
discussed here, can also affect the fragmentation of the flow. Bromm and Loeb
(2003) have shown that isolated $2\sigma$ perturbations in the presence of
metal-free baryons lead to the formation of a single massive clump at the center.
When $H_2$ cooling is suppressed, these clumps enter the isothermal collapse stage
without further fragmentation. Dijkstra et al. (2008) argue that only a small
fraction of high-$z$ halos will be subject to a strong ionizing UV continuum
(so-called Lyman-Werner background) capable of dissociating the $H_2$ molecules
and suppressing the molecular cooling and, therefore, the gas fragmentation.
Furthermore, Population~III stars can enrich the interstellar and intergalactic
medium with metals that will allow the cooling to proceed below a few$\times 10^3$K
even in the absence of $H_2$ (Begelman et al. 2006; Wise \& Abel 2008, 
and references therein). The focal point of this Letter is that molecular and 
metal cooling is not the determining factor in the fragmentation of rapid
inflows associated with the galaxy formation. 

The first hints that fragmentation may be avoidable in bars-in-bars collapse have 
been provided recently by numerical simulations, which reveal that nested bar 
instabilities accompany the gravitational collapse, as predicted. 
The flow encompasses two different dynamical regimes: (1) a dark matter 
(DM)-dominated regime, from the growing halo virial 
radius down to a few pc; and (2) a gas-dominated regime, within the central few pc.
The latter regime is characterized by the dynamical decoupling of
self-gravitating gas from the DM background (Englmaier \& Shlosman 2004).
The characteristic radius of a few pc applies to DM halos with virial temperatures
somewhat in excess of $10^4$~K. More massive halos will have correspondingly
larger gas-dominated central regions, and hence the final product of gravitational
collapse can be much more massive.

It is not surprising that supersonically inflowing gas should become turbulent but 
somewhat surprising that the turbulence should maintain a virial level throughout 
the flow, given its high dissipation rate --- this requires an efficient driving
mechanism. On larger scales, the turbulence can be 
driven by a variety of mechanisms inherent in the hierarchical buildup of structure 
in the Universe, such as mergers. On smaller scales, the turbulence can be driven 
by the gravitational energy released during the infall. As gas with some
seed $J$ goes into gravitational collapse, the centrifugal barrier will
slow it down at some cylindrical radius $R$. This will happen 
concurrently with the decay of turbulence
and will drive $T/|W|$ over the instability threshold. The instability develops 
on a dynamical 
timescale. The growth of the barlike mode in the 
gas will excite shocks and force the gas to lose rotational support, forcing
it into further gravitational collapse and pumping energy into turbulent motions. 
Thus, we suggest that the persistence of virial, supersonic turbulence in a 
self-gravitating inflow is the {\it consequence} of a self-regulating instability.  
This cascade will be quenched only when the gas becomes optically thick, resulting
in the sharp rise of thermal energy compared to turbulent energy. We quantify the
above processes in \S3.

As an example, we assume a time-independent, self-similar, self-gravitating, 
isothermal disklike configuration with Keplerian velocity $v_{\rm K}$ independent 
of radius. (The latter is a natural outcome for a non-rotating flow [Larson 1969], 
but it can be extended to inflows with substantial rotation.)
The surface density scales as $R^{-1}$.  If the inflow speed is roughly 
independent of radius 
(as suggested by the simulations), it does not affect the radial momentum equation 
(since $v_{\rm R} {\rm d}v_{\rm R}/{\rm d}R \approx 0$).  One can then construct 
a crude model 
for the radial structure of the inflow.  Neglecting thermal pressure, one finds a 
tradeoff between the typical turbulent velocity $v_{\rm t}$ and the rotation speed 
$v_\phi$, $v_\phi^2 = v_{\rm K}^2 - 2 v_{\rm t}^2$. If $v_{\rm t}$ is small, the 
rotation is close to Keplerian and the ratio $T/|W|$, which is 
roughly proportional to $(v_\phi/v_{\rm K})^2$, exceeds the threshold for bar 
instability. As the radial inflow slows down, the turbulence decays  --- to the 
point where $T/|W|$ approaches the stability threshold.   
The decreasing ratio $v_{\rm t}/v_\phi$ will lead to self-similar vertical 
structure as well, with 
$\rho\propto R^{-2}$, as seen in the simulations. (We note that the self-similar, 
self-gravitating disk models of Toomre [1982] provide a promising platform for 
refining this picture.) 
This example also holds, with straightforward quantitative modifications, when the 
gas inflow proceeds in the background potential of the DM halo. 
 
\section{Damping fragmentation in thick disks}

\begin{figure*}[ht!!!!!!!!!]
\centering
   \includegraphics[angle=0,scale=0.275]{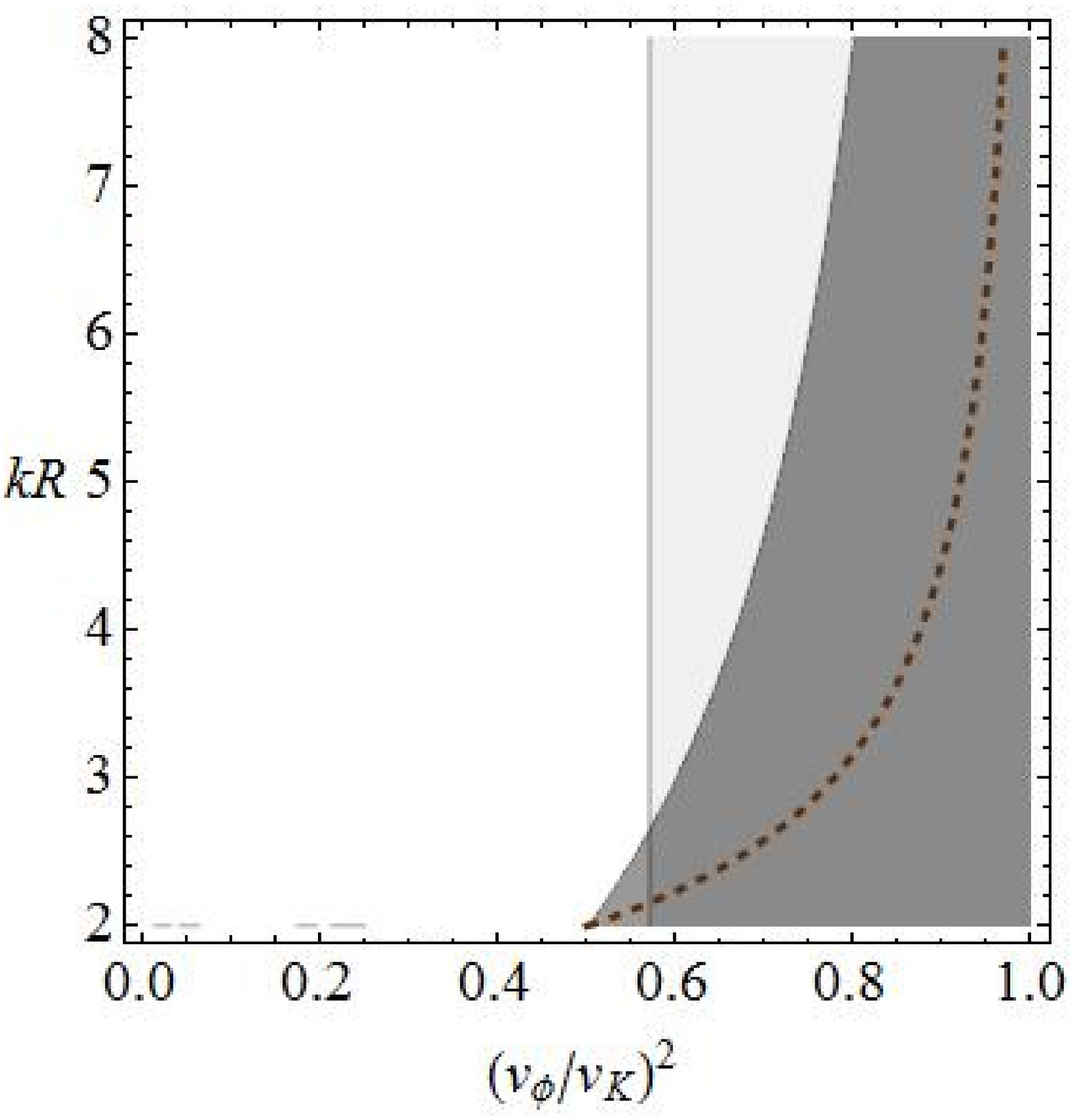}
   \includegraphics[angle=0,scale=0.25]{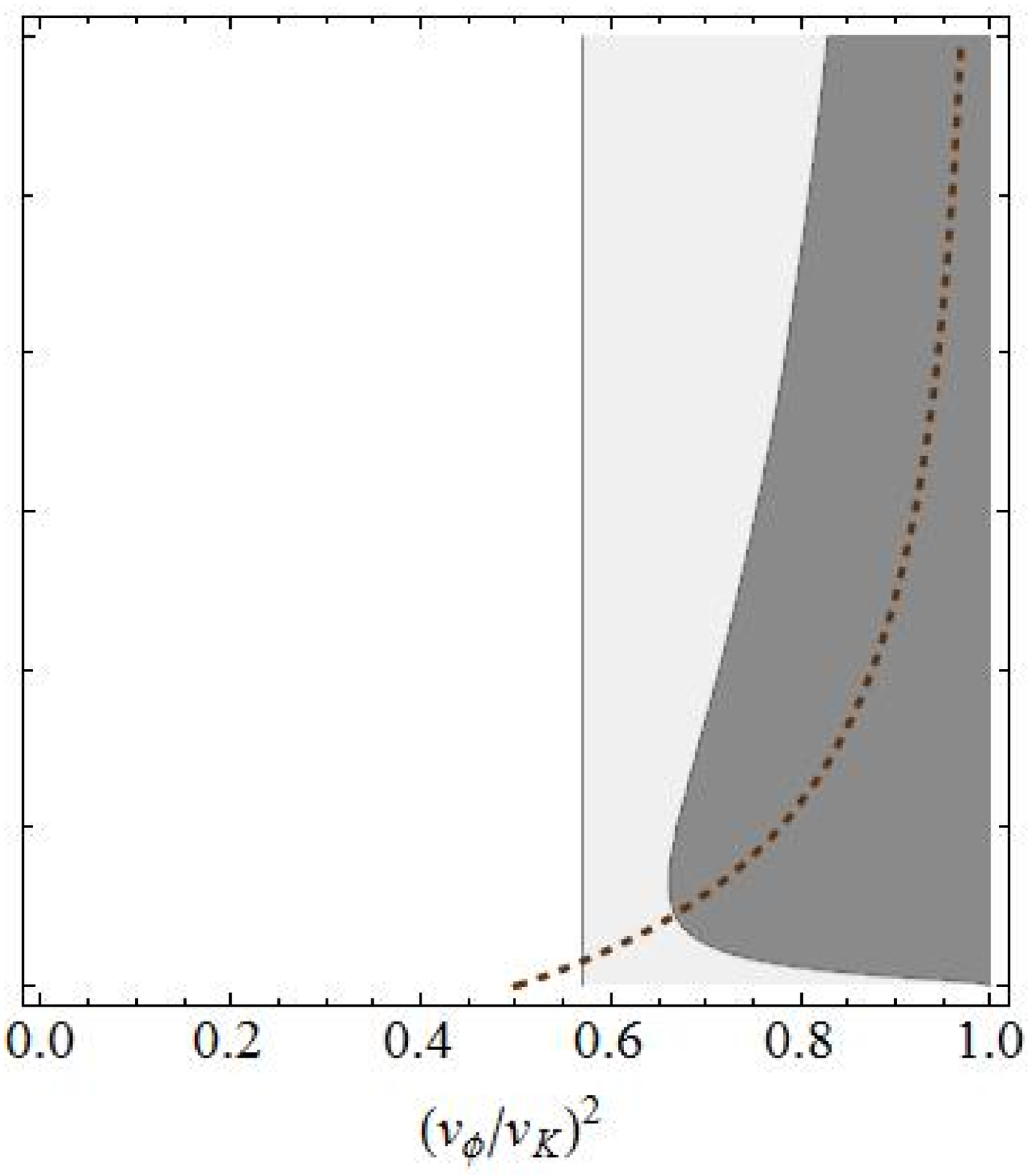}
   \includegraphics[angle=0,scale=0.25]{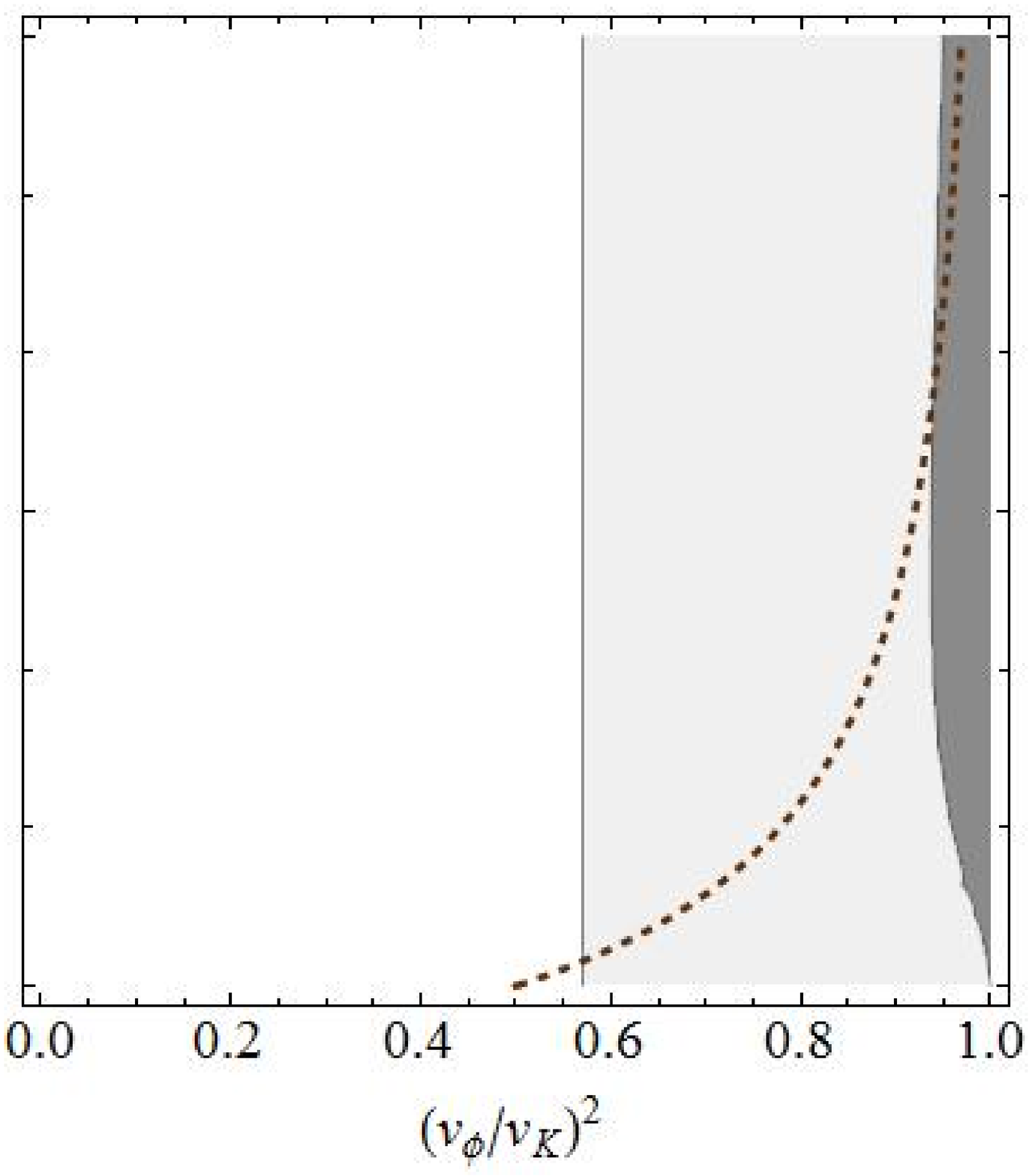}
  \caption{Stability of thin and thick disks to fragmentation and to bar
     formation in the $x\equiv (v_{\phi}/v_{\rm K})^2$ --- $y\equiv
     kR$ plane. The dark-shaded regions show where
     the disk is unstable to fragmentation, while the vertical line represents
     the boundary to whose right the disk becomes bar unstable. The left
     frame describes a razor-thin, 2-D disk; the middle frame corresponds to the
     case where the disk thickness is only $5\%$ of the actual thickness; while
     the third frame corresponds to the actual thickness.  $kh > 1$ above the
     dashed curve in each panel. }
   \label{stabfig}
\end{figure*}

The next aspect of the flow that needs to be addressed is fragmentation.  If 
$v_{\rm t} \ll v_\phi$ and the thermal energy dominates over the turbulent one, 
one can show that the disk is strongly unstable to breakup into rings (Toomre 
1964).  A direct application of the Toomre criterion, which is a generalized
Jeans instability applied to razor-thin, differentially-rotating disks, 
shows that the disk becomes 
more stable as $v_{\rm t}$ increases at the expense of $v_\phi$. 
We test whether a self-gravitating gaseous disk which is
subject to the (global) bar instability is stable locally, i.e., whether it is 
subject to fragmentation. Here, we estimate the effect of the finite thickness 
induced by the turbulence
on the local stability of the disk. In the following, we refer to 2-D disks as 
`thin' disks, and to disks with a finite thickness as `thick' disks. The latter 
are fully 3-D rotating configurations which are supported vertically (along the 
rotation axis) against gravity by turbulent motions, while in the equatorial 
plane, both rotation and turbulent motions contribute to this support.

The dispersion relation for a razor-thin, 2-D gaseous disk with a surface density
$\Sigma$ and a turbulent velocity $v_{\rm t}$ (which replaces the sound speed) is 
given by 
\begin{equation}
\omega^2=\kappa^2+v_{\rm t}^2k^2-2\pi G\Sigma |k| ,
\end{equation}
where $\kappa$ is the epicyclic frequency, $k$
is the radial wavevector and $\omega(k)$ is the wave frequency in the inertial frame. 
The isothermal disks considered here have a flat rotation
curve and consequently $\kappa^2=2\Omega^2$, where $\Omega$ is the angular velocity in
the disk. In an unperturbed self-gravitating disk, the local Keplerian velocity in the
disk is given by $v_{\rm K}^2=v_\phi^2+2v_{\rm t}^2$, where $v_\phi=\Omega R$. The
tradeoff between the turbulent ``pressure" and the rotational support against gravity,
for a fixed $v_{\rm K}$, is crucial for understanding the disk dynamics and its
stability.

Thin gaseous disks are stable to fragmentation if the Toomre parameter $Q$
satisfies
\begin{equation}
Q= \frac{\kappa v_{\rm t}}{\pi G\Sigma } > 1.
\end{equation}
These disks are stabilized by rotation on larger scales and by the thermal (or turbulent)
pressure on smaller scales. Sufficiently large $Q$ ensures that these two scales overlap
and the disk does not fragment. Both thermal and turbulent pressures in these disks are
assumed to act in the equatorial plane only.   

In reality, a disk with finite,
isotropic pressure will have a finite vertical thickness, $h$, which is
maintained by the turbulence: $h=(v_{\rm t}/v_{\rm K})R$. The finite disk thickness
affects the dispersion relation (eq.~1) by decreasing the value of the destabilizing
gravity term (which appears with the negative sign), as follows:
\begin{equation}
\omega^2=\kappa^2+v_{\rm t}^2k^2-\frac{2\pi G\Sigma |k|}{1+|k|h} .
\end{equation}
The correction factor, $(1+|k|h)^{-1}$, has been obtained by solving the Poisson equation
along the rotation ($z$) axis, assuming that the disk has an exponential density profile,
$\Sigma(z)\sim e^{-|z|/h}$.  Note that eq.~(3) reduces to the usual dispersion relation for
Jeans instability in the limit $h/R\ll 1$. The mass enclosed within a radius $R$ by an
isothermal, self-gravitating disk is given by $M=2\pi\Sigma R^2$, so that by using
$GM/R^2=v_{\rm K}^2/R$, the dispersion relation (eq.~3) can be re-written as
\begin{equation}
\frac{\omega^2R^2}{v_{\rm K}^2}=2\frac{v_\phi^2}{v_{\rm K}^2}+\frac{v_{\rm t}^2}
     {v_{\rm K}^2}(kR)^2-\frac{kR}{1+(kR)\frac{v_{\rm t}}{v_{\rm K}}} .
\end{equation}
Introducing dimensionless variables, $y\equiv kR$, $x\equiv
(v_\phi/v_{\rm K})^2$ and $\zeta\equiv \omega R/v_{\rm K}$, so that
$(v_{\rm t}/v_{\rm K})^2=(1-x)/2$, we get
\begin{equation}
\zeta^2=2x + \frac{1-x}{2}y^2 -\frac{y}{1+y(\frac{1-x}{2})^{1/2}} ,
\end{equation}
which is the dispersion relation for self-gravitating, isothermal gaseous disks with 
a finite thickness contributed by turbulent motions. Our goal is to obtain the solution 
of this equation at the threshold for bar instability in such disks. The latter is 
given by Christodoulou et al. (1995) as
\begin{equation}
\alpha\equiv\bigg(\frac{f}{2}\frac{T}{|W|}\bigg)^{1/2} > \alpha_{\rm crit}
    \approx 0.35 ,
\end{equation}
where $T=(v_\phi/v_{\rm K})^2(v_{\rm K}^4R/2G)=x(v_{\rm K}^4R/2G)$ is the kinetic 
energy of
the disk out to radius $R$, $|W|=\int_0^R GM{\rm d}M/R=v_{\rm K}^4R/G$ is the potential
energy contained within the same radius, and $f$ is the shape factor.  In our 
dimensionless
units, $T/|W|=x/2$.  The shape factor is $f=1$ for a razor-thin disk and 
$f=(2/3)(1+T/|W|)$
for thick, differentially rotating spheroidal mass configurations.
At marginal global stability of the disk, $\alpha=\alpha_{\rm crit}$; by substituting
$\alpha_{\rm crit}$ into eq.~(6), we obtain $x_{\rm crit}\approx 0.57$.
 
Fig.~1 displays the local and global stability regimes in the $x-y$, i.e.,
$(v_{\phi}/v_{\rm K})^2-kR$, plane under imposed conditions as described below. The 
dark-grey regions delineate regimes of local instability, i.e., that are subject 
to 
fragmentation. In each frame, gas that lies in the left part of the diagram has 
negligible rotation, which corresponds to the usual Jeans instability. The right part 
of each frame shows rotating turbulent gas, corresponding rather to Toomre's 
instability. The left frame depicts stability conditions under the artificial 
assumption that the disk is 2-D (razor-thin).  As expected, the right hand-side, for 
$x>0.49$, is unstable over a wide range of wavelengths, because the turbulent 
pressure and rotation are not sufficient to prevent fragmentation.  The second
frame corresponds to a case in which we have (artificially) allowed the disk to 
maintain 5\% of its thickness compared to what it should actually have based on 
$v_{\rm t}$. The unstable region has moved to $x>0.66$.  When the full value of disk 
thickness is accounted for (right frame), the stability boundary lies at 
$x\geq 0.94$, i.e., in the regime of rapid rotation.
 
The light gray area shows the region of global (i.e., bar) instability calculated 
above, $x>x_{\rm crit}\approx 0.57$. We observe that gas flows that possess fully 
developed turbulence must move from left to right in Fig.~1, as the turbulence 
decays with time. Hence the gas will first reach the threshold for bar instability 
while it is still locally stable. 
When the finite thickness of the disk is 
accounted for in the dispersion relation, the global (i.e., bar) instability sets 
in well before the disk is able to fragment.

What about fragmentation on small (sub-Jeans) scales, due to the compression of gas 
in the isothermal shocks that characterize supersonic turbulence?  To address this, 
we adopt a widely used model of supersonic turbulence (e.g., Padoan 1995; Krumholz 
\& McKee 2005; 
for reviews: Mac~Low \& Klessen 2004; Elmegreen \& Scalo 2004; McKee \& Ostriker 
2007).  We estimate the fraction of the material swept up by the shocks that becomes 
Jeans unstable, by following the formalism of Padoan \& Nordlund (2002) and Krumholz 
\& McKee (2005).  The turbulence is characterized by a lognormal probability 
distribution, 

\begin{equation}
p(x)=\frac{1}{(2\pi\sigma_{\rm p}^2)^{0.5}} \frac{1}{x} {\rm exp}\bigg[-\frac{({\rm 
    ln}\,x-\overline{{\rm ln}\,x})^2}{2\sigma_{\rm p}^2}\bigg] , 
\end{equation}
with the mean $\overline{{\rm ln}\,x} = 
-0.5\sigma_{\rm p}^2$, where $x\equiv \rho/\rho_0$ and $\rho_0$ is the mean density, 
and the dispersion $\sigma_{\rm p}\approx [{\rm ln}\,(1+3\mathcal{M}^2/4)]^{0.5}$. 

Two spatial scales characterize such turbulence: the Jeans scale, $\lambda_{\rm J}=
(\pi c_{\rm s}^2/G\rho)^{0.5}$, and the transition scale, $\lambda_{\rm s}$, below 
which the flow becomes subsonic. The latter scale relates to the supersonic velocity 
dispersion, $\sigma_{\rm t}$, by  $\sigma_{\rm t}\sim c_{\rm s} 
(R/\lambda_{\rm s})^{\beta}$, which is valid over a large dynamic range; with 
$\beta\sim 
0.4-0.5$ (Larson 1981; Krumholtz \& McKee 2005). In the overdense regions 
$\lambda_{\rm J}\sim \lambda_{\rm J0}x^{-0.5}$, where 
$\lambda_{\rm J0}$ is the Jeans length at $\rho_0$. To estimate the fraction of 
swept up gas that is Jeans-unstable {\it over one free-fall time}, $f_{\rm J}$, 
we integrate the lognormal distribution from  
$x_{\rm crit}\equiv \rho_{\rm crit}/\rho_o =  (\lambda_{\rm J0}/\lambda_{\rm s})^2 
\approx 6\gamma_{0.5}^2\mathcal{M}^2$ to infinity, where we have assumed $\beta=0.5$
and $\gamma_{0.5}\equiv 0.5v_{\rm t}/v_{\rm K}\sim 1$. 
Strikingly, the dependence of $f_{\rm J}$ on  $\mathcal{M}$ assures that 
$f_{\rm J}\ltorder 2\times 10^{-2}$ for $\mathcal{M}\gtorder 3$ --- such Mach
numbers are routinely observed in simulations of inflows associated with galaxy
formation. This differs from the gas evolution in 
galactic disks, where supersonic turbulence develops as well, but the turbulent 
velocities are very sub-virial (because of lack of sufficient driving).
Further high-resolution numerical simulations are needed to shed light on the 
intricacies of shock dissipation in supersonic turbulence over a wider range of
Mach numbers.

\section{Discussion}

To summarize, we have argued that the hypersonic collapse of self-gravitating 
gas driven by nested gaseous bars is, counterintuitively, {\it not} susceptible 
to fragmentation.  Fragmentation is suppressed whenever the gas temperature 
remains below the virial temperature; moreover, the 
resistance to fragmentation increases with the infall Mach number.  
If the gas has sufficient time to achieve a quasi-equilibrium, when it has
cooled down or its turbulence has decayed, it will be subject to 
fragmentation. If, however, a decrease in the level of turbulence
leads to renewed non-axisymmetric instability, as we have argued, then the 
infall will recommence.
The slowdown of radial inflow in regions of relative 
stability leads to spikes at which $\alpha$ substantially exceeds 
$\alpha_{\rm crit}$, leading to the creation of a distinct bar and a rapid 
acceleration of infall.  Thus, the bars-in-bars mechanism
seems to be a highly intermittent one, as originally envisaged (Shlosman et al. 
1989, 1990).

If confirmed by carefully designed numerical experiments, this scenario could have 
important implications for the formation and evolution of the central regions of 
galaxies and, in particular, the formation of supermassive black holes (SBHs).  
It was suggested that the formation of SBH seeds by direct 
collapse would be suppressed in $10^4$ K halos that were metal-enriched or had 
significant cooling by H$_2$, because the infalling gas would fragment and form 
stars (e.g., Bromm \& Loeb 2003; Begelman et al. 2006; Wise \& Abel 2008). If this 
is not true, then a much larger fraction of $10^4$ K halos could 
form single massive objects. We note that, within this framework, DM halos with 
larger virial temperatures could form more massive SBH seeds via direct collapse, 
if a sufficient amount of gas accumulates in their midst.
Moreover, this mechanism would open the door to the formation or rapid 
growth of SBHs at later epochs in such massive halos, for example, following 
gas-rich mergers.  There are also implications for the triggering of 
starbursts and the fueling of AGN in galactic nuclei where an SBH 
already exists.

\acknowledgements
We thank Phil Armitage, Bruce Elmegreen, Martin Haehnelt, Andrew Hamilton and John 
Wise for very helpful discussions. Support by the NSF, NASA ATP and LTSA 
programs, and STScI is gratefully acknowledged. STScI is operated by AURA, Inc., 
under NASA contract NAS 5-26555. I.S. thanks the JILA Fellows for support.

%----------------------------------------------------------------------
%----------------------------------------------------------------------
%----------------------------------------------------------------------
%----------------------------------------------------------------------


\begin{thebibliography}{}

\bibitem[]{}Begelman, M.C., Volonteri, M. \& Rees, M.J. 2006, \mnras, 370, 289

\bibitem[]{}Blandford, R.D. \& Payne, D.G. 1982, \mnras, 199, 883

\bibitem[]{}Bromm, V. \& Loeb, A. 2003, \apj, 596, 34

\bibitem[]{}Chandrasekhar, S. 1969, Ellipsoidal Figures of Equilibrium, New
    Haven: Yale Univ. Press

\bibitem[]{}Christodoulou, D.M., Shlosman, I. \& Tohline, J.E. 1995,
    \apj, 443, 551

\bibitem[]{}Dijkstra, M., Haiman, Z., Mesinger, A. \& Wuithe, J.S.B. 2008, \mnras, 
    391, 1961

\bibitem[]{}Elmegreen, B.G. \& Scalo, J. 2004, ARAA, 42, 211

\bibitem[]{}Englmaier, P. \& Shlosman, I. 2004, \apj, 617, L115

\bibitem[]{}Heller, C.H., Shlosman, I. \& Athanassoula, E. 2007, \apj, 657, L65

\bibitem[]{}Krumholz, M.R. \& McKee, C.F. 2005, \apj, 630, 250

\bibitem[]{}Larson, R.B. 1969, \mnras, 145, 271

\bibitem[]{}Larson, R.B. 1981, \mnras, 194, 826

\bibitem[]{}Levine, R., Gnedin, N.Y., Hamilton, A.J.S. \& Kravtsov, A.V. 2008,
      \apj, 678, 154

\bibitem[]{}Lynden-Bell, D. \& Kalnajs, A.J. 1972, \mnras, 157, 1

\bibitem[]{}Mac Low, M.-M. \& Klessen, R.S. 2004, Rev. Mod. Phys, 76, 125

\bibitem[]{}McKee, C.F. \& Ostriker, E.C. 2007, ARAA, 45, 565

\bibitem[]{}Paczy\'nski, B. 1978, Acts Astron., 28, 91

\bibitem[]{}Padoan, P. 1995, \mnras, 277, 377

\bibitem[]{}Padoan, P. \& Nordlund, A. 2002, \apj, 576, 870

\bibitem[]{}Regan, J.A. \& Haehnelt, M.G. 2009, \mnras, 396, 343 

\bibitem[]{}Romano-Diaz, E., Shlosman, I., Heller, C.H. \& Hoffman, Y. 2008, \apj,
     687, L13

\bibitem[]{}Shibata, M., Baumgarte, T.W. \& Shapiro, S.L. 2000, \apj, 542, 453

\bibitem[]{}Shlosman, I., Begelman, M.C., Frank, J. 1990, Nature, 345, 679

\bibitem[]{}Shlosman, I., Frank, J. \& Begelman, M.C. 1989, Nature, 338, 45

\bibitem[]{}Toomre, A. 1964, \apj, 139, 1217

\bibitem[]{}Toomre, A. 1982, \apj, 259, 535

\bibitem[]{}Wise, J.H. \& Abel, T. 2008, \apj, 685, 40

\bibitem[]{}Wise, J.H., Turk, M.J. \& Abel, T. 2008, \apj, 682, 745


\end{thebibliography}
\end{document}